\documentclass[fleqn,usenatbib]{mnras}

\usepackage{newtxtext}

\usepackage{ae,aecompl}

\usepackage{graphicx}	
\usepackage{amsmath}	
\usepackage{amssymb}	
\usepackage[dvipsnames]{xcolor} 

\newcommand{\Msun}{ h^{-1}{\rm M_{ \odot}}}
\newcommand{\hMpc}{ h^{-1}{\rm Mpc}}

\newcommand{\ihMpc}{ h\,{\rm Mpc}^{-1}}

\title[Matter bispectrum with baryons]{Simultaneous modelling of matter power spectrum and bispectrum in the presence of baryons}

\author[G. Aric\`o et al.]{Giovanni Aric\`o$^{1}$\thanks{E-mail:giovanni\_arico001@ehu.eus (GA)},
Raul E. Angulo$^{1,2}$,
Carlos Hern\'andez-Monteagudo $^{3,4,5}$
\newauthor
Sergio Contreras$^{1}$, \& Matteo Zennaro$^{1}$.
\\
\\
$^{1}$Donostia International Physics Center (DIPC), Paseo Manuel de Lardizabal, 4, 20018, Donostia-San Sebasti\'an, Guipuzkoa, Spain.\\
$^{2}$IKERBASQUE, Basque Foundation for Science, 48013, Bilbao, Spain.\\
$^{3}$Centro de Estudios de F\'isica del Cosmos de Arag\'on, Unidad Asociada CSIC, Plaza San Juan 1, 44001 Teruel, Spain.\\
$^{4}$Instituto de Astrof\'isica de Canarias, Spain.\\
$^{5}$ University of La Laguna, Spain.
}

\date{Accepted XXX. Received YYY; in original form ZZZ}

\pubyear{2020}
\hypersetup{draft=False}
\begin{document}
\label{firstpage}
\pagerange{\pageref{firstpage}--\pageref{lastpage}}
\maketitle

\begin{abstract}

We demonstrate that baryonification algorithms, which displace particles in gravity-only simulations according to physically-motivated prescriptions, can simultaneously capture the impact of baryonic physics on the 2 and 3-point statistics of matter. Specifically, we show that our implementation of a baryonification algorithm jointly fits the changes induced by baryons on the power spectrum and equilateral bispectrum on scales up to $ k = 5 \ihMpc$ and redshifts $0 \le z \le 2$, as measured in six different cosmological hydrodynamical simulations. The accuracy of our fits are typically $\sim 1\%$ for the power spectrum, and for the equilateral and squeezed bispectra, which somewhat degrades to $\sim 3\%$ for simulations with extreme feedback prescriptions. Our results support the physical assumptions underlying baryonification approaches, and encourage their use in interpreting weak gravitational lensing and other cosmological observables.
\end{abstract}

\begin{keywords}
 large-scale structure of Universe -- cosmological parameters -- cosmology: theory
\end{keywords}



\section{Introduction}
\label{sec:into}

Despite large efforts of the scientific community, the nature of dark energy and dark matter remains elusive. Even if the standard  $\Lambda$CDM model has successfully passed many independent tests in the last decades, recent tensions in the estimated values of the Hubble constant and in the amplitude of the linear fluctuation have been pointed out as a possible window to physics beyond $\Lambda$CDM \citep[e.g.][]{Verde2019,Wong2020}.
To successfully solve these tensions, it is paramount that current and upcoming cosmological surveys extract the maximum amount of cosmological information at the low redshifts, where dark energy and dark matter are more accessible \citep[][]{Planck2018,DES12018,jpas,euclid,DESI,HSC_SSP}. For many observables and statistics, the limiting factor will be the predictability and accuracy of theoretical models employed to analyse the data.

For the case of next-generation weak lensing surveys, the largest theoretical uncertainty is given by baryonic physics -- gas cooling, star formation, and feedback, for instance, modify significantly the total mass distribution in the universe in a way that is not possible to accurately predict from first principles. On the other hand, if these baryonic processes are modelled appropriately, then we could extract more cosmological information, and possibly also constrain astrophysical processes.

A promising approach to incorporate the baryonic effects in models for the cosmic density field is {\it baryonification} \citep{Schneider&Teyssier2015, Schneider&Teyssier2019, Arico2020}. Briefly, these algorithms, a.k.a. {\it baryon correction models} (BCM), displace particles in gravity-only simulations according to physically-motivated recipes designed to mimic the effects produced by baryons in the Large Scale Structure (LSS) of the universe. This method has been extensively tested against many hydrodynamical simulations, and it is shown to be very accurate in capturing the changes induced by baryons on the power spectrum \citep{Schneider2020,Arico2020}.

In general, baryons are expected to modify the full density field and thus the whole hierarchy of $N$-point functions, not only the power spectrum. Indeed, \cite{Foreman2019}  recently showed that baryonic effects on the bispectrum of hydrodynamical simulations are present, and that they carry extra information with respect to the power spectrum. Given the simplifications and assumptions of baryonification methods -- e.g. spherically symmetric displacements, dependences on halo mass, and neglected physical processes --  it is unclear whether they would be able to consistently model baryonic effects on the power spectrum and bispectrum.

Exploring the predictions of baryonification for the bispectrum is an important topic since it could highlight the pitfalls of the method or, instead, could support the correctness of the whole approach. Additionally, this comparison would represent an independent test of the method since higher-order statistics were never employed in the formulation of the baryonification algorithm.

Motivated by these findings, in this paper we extend the analysis of the baryonification
algorithm presented in \cite{Arico2020} (hearafter A20), and here revisited, to the bispectrum. First, we extend the model to account for gas that has been reaccreted by halos, and then show how different model parameters change the power spectrum and bispectrum. Then, we show that our baryonification implementation can simultaneously reproduce, to better than $3\%$, the power spectrum and bispectrum measured in six state-of-the-art hydrodynamical simulations at $k \le 5\ihMpc$ and at $z \le 2$. We furthermore explore the impact that the different components of the model, e.g. central galaxy, ejected gas and back-reaction onto dark matter, have on the matter bispectrum.

This paper is structured as follow: in \S\ref{sec:sim} we describe our numerical simulations, while in \S\ref{sec:mbcm} we present our methodologies for baryonic and cosmology modelling of the density field. In \S\ref{sec:impact} we discuss the impact of baryons in the bispectrum, whereas in \S\ref{sec:hydro_bcm} we show our fits to the hydrodinamical simulations. We give our conclusions in \S\ref{sec:conclusions}.

\begin{figure*}
  \includegraphics[width=.9\linewidth]{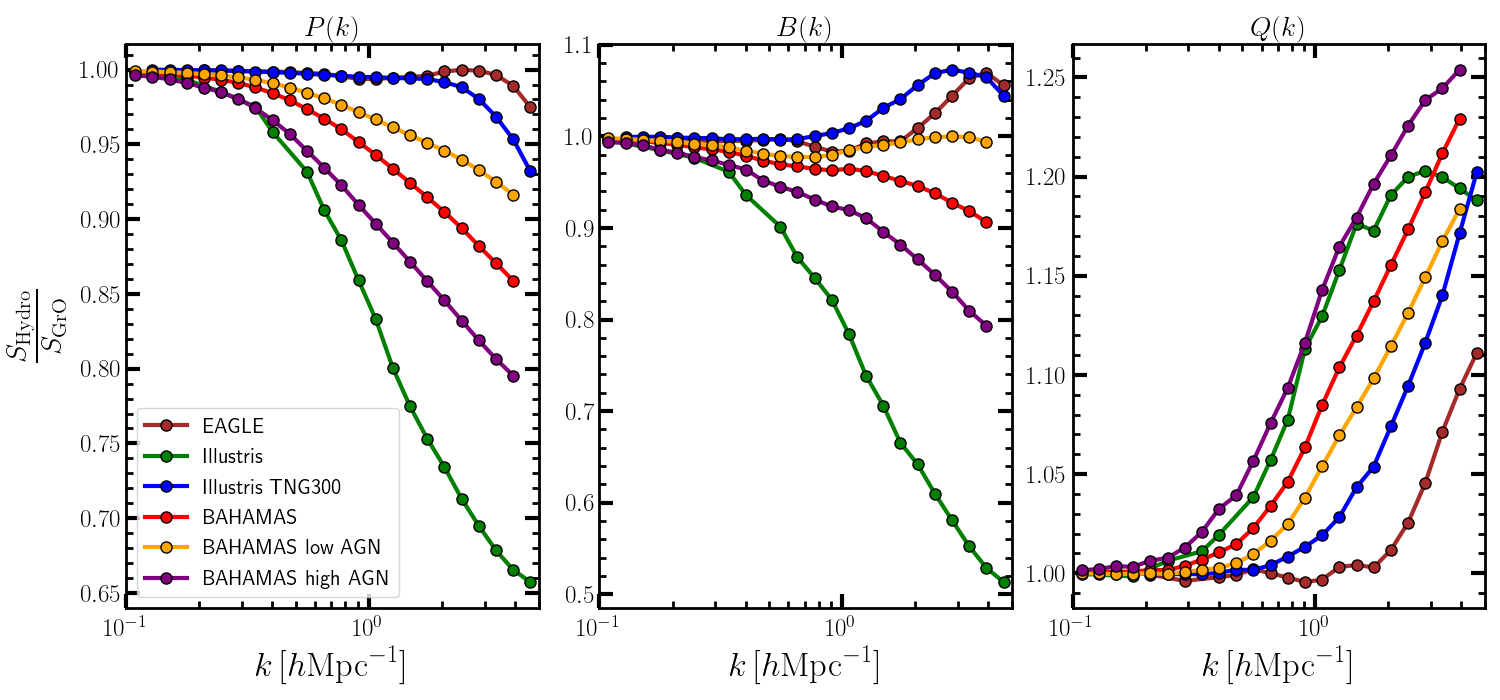}
  \caption{Baryonic effects at $z=0$ on the power spectrum (left panel), equilateral bispectrum (central panel) and reduced bispectrum (right panel), measured in 6 hydrodynamical simulations: BAHAMAS (standard, low and high AGN), EAGLE, Illustris and Ilustris TNG-300. We display the ratio of $S = \{P, B, Q\}$ estimated in the full hydrodynamical simulation to that in their respective gravity-only counterpart.}
  \label{fig:data}
\end{figure*}

\section{Numerical simulations}
\label{sec:sim}

In this work we use the same suite of $N$-body simulations used in A20. We refer the reader to it for further details, and here we only provide a brief description.

Our gravity-only simulations were carried out with \texttt{l-gadget-3} \citep{Angulo2012}, a modified version of \texttt{gadget} \citep{Springel2005GADGET}. We employ simulations of box sizes: L=64, 128, and 256 $\hMpc$ containing $192^3$, $384^3$, $768^3$ particles, respectively. We adopt the {\it Nenya} cosmology, as defined by \cite{Contreras2020}:
 $\Omega_{\rm cdm}=0.265$, $\Omega_{\rm b}=0.050$, $\Omega_{\Lambda}=0.685$, $H_0=60$ ${\rm km}\, s^{-1}\,{\rm Mpc}^{-1}$,  $n_s=1.010$, $\sigma_8=0.90$, $\tau=0.0952$, $\sum m_{\nu}=0$, $w_0=-1$, and $w_a=0$.
These parameters are optimal to rescale them to a large range of cosmologies \citep{Contreras2020, Angulo2020}.
To test the accuracy of this rescaling, we will consider two additional simulations of $512 \,\hMpc$ and $1536^3$ particles: one adopting the {\it Nenya} cosmology, and the other a massless neutrino {\it Planck} cosmology \citep{Planck2018}
 \footnote{$\Omega_{\rm cdm}=0.261$, $\Omega_{\rm b}=0.049$, $\Omega_{\Lambda}=0.699$, $H_0=67.66$ $ {\rm km}\, s^{-1}\,{\rm Mpc}^{-1}$,  $n_s=0.966$, $\sigma_8=0.81$, $\tau=0.0561$, $\sum m_{\nu}=0$, $w_0=-1$, $w_a=0$.}.

The initial conditions of all our simulations were computed with the ``fixed and paired'' technique described in \cite{Angulo&Pontzen2016}, thus their cosmic variance is heavily suppressed.
To compute the statistics of the density field, we use catalogue of simulation particles, selected homogeneously, diluted by a factor of $4^3$. If not specified otherwise, our results will be computed with our L=256 $\hMpc$ simulation, with which we expect our results to be converged to about 2\% for both bispectrum and power spectrum, as Appendix~\ref{app:convergence} shows.

\begin{figure*}
  \includegraphics[width=0.9\linewidth]{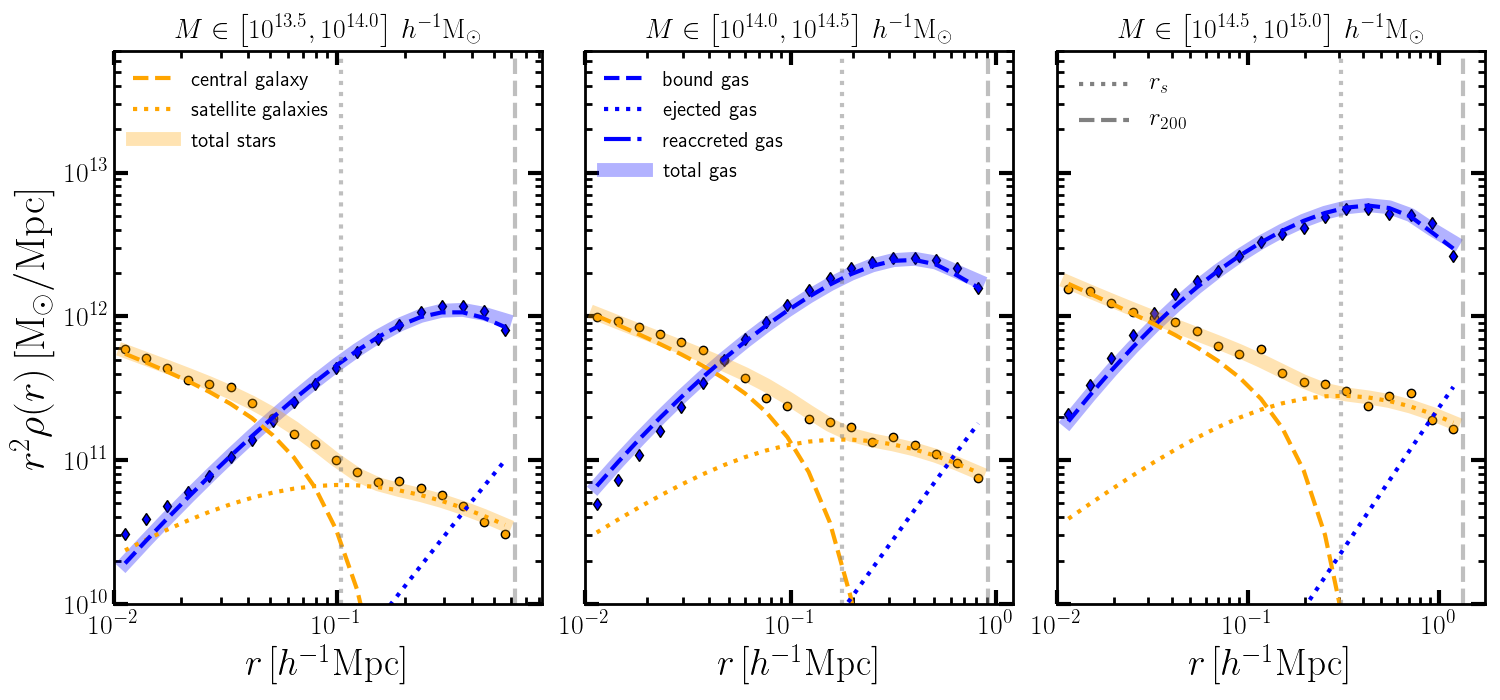}
  \caption{Density profiles of gas (blue diamonds) and stars (orange circles) as measured in the Illustris TNG-300 simulation, at $z=0$. Each panel shows a different mass bin: $10^{13.5}-10^{14} \, \Msun$ (left panel), $10^{14}-10^{14.5} \, \Msun$ (central panel), and $10^{14.5}-10^{15} \, \Msun$ (right panel). The baryonification model that best fits simultaneously the three density profiles is shown as blue and orange shaded bands for gas and stars, respectively. The different gas and stellar subcomponents are displayed with different line styles, according to the legend. Note that the reaccreted gas density is consistent with zero, thus not appearing in the plot.}
  \label{fig:density_profiles}
\end{figure*}

\subsection{Measurement of power spectra and bispectra}
Considering an overdensity field in Fourier space $\delta( {\bf k} )$, we define the power spectrum as

\begin{equation}
\langle \delta({\bf k_1}) \delta({\bf k_2}) \rangle \equiv (2\pi)^3 \delta^D({\bf k_1} + {\bf k_2}) P(k_1)
\label{eq:pk}
\end{equation}
and the bispectrum as

\begin{equation}
 \langle \delta( {\bf k_1} ) \delta({\bf k_2}) \delta({\bf k_3}) \rangle \equiv (2\pi)^3 \delta^D({\bf k_1}+{\bf k_2}+ {\bf k_3})  B(k_1,k_2,k_3),
\label{eq:bk}
\end{equation}

\noindent where $\langle...\rangle$ denotes the ensemble average and $\delta^D$ is the Dirac's delta. To reduce the dependence of the bispectrum on the power spectrum and cosmology, we will mostly consider the {\it reduced bispectrum}  \citep{Scoccimarro2000,Sefusatti2007}, defined as

\begin{equation}
Q(k_1,k_2,k_3) \equiv \frac{B(k_1,k_2,k_3)}{P(k_1)P(k_2)+P(k_2)P(k_3)+P(k_1)P(k_3)}.
\label{eq:qk}
\end{equation}

We will mostly focus on the equilateral configuration, $k_1 = k_2 =k_3$, since it is expected to contain the most independent information from the power spectrum. In this case, Eq.~\ref{eq:qk} is reduced to:

\begin{equation}
Q(\texttt{k}) = \frac{B(\texttt{k})}{3P(k)^2}.
\label{eq:qk_equilateral}
\end{equation}

We measure the bispectrum using \texttt{bskit} \citep{Foreman2019}\footnote{\url{https://github.com/sjforeman/bskit}}, an extension of \texttt{nbodykit} \citep{Hand2018} which uses a Fast Fourier Transform (FFT)-based bispectrum estimator \citep{Scoccimarro2000}. Both the bispectrum and the power spectrum are measured in two interlaced grids \citep{Sefusatti2016} employing a {\it triangular shaped cloud} mass assignment scheme. The shot noise contribution is estimated as $1/\bar{n}$ for the power spectrum, and as $1/\bar{n}^2+1/\bar{n} [P(k_1)+P(k_2)+P(k_3)]$ for the bispectrum, and subtracted. Finally, we have rebinned all the measurements in 25 logarithmic bins over the interval $[0.1,5.0] \, \ihMpc$. Additionally, when measuring the clustering on small scales, we use the ``folding''
technique \citep{Jenkins1998,Colombi2009}, described in Appendix \ref{app:fold}, which reduces CPU and memory usage.

We will compare our results against the power spectra and bispectra from a number of cosmological hydrodynamical simulations, as measured by \cite{Foreman2019}\footnote{\url{https://github.com/sjforeman/hydro_bispectrum}}.
Specifically, we use four state-of-the-art hydrodynamical simulations: BAHAMAS \footnote{\url{http://www.astro.ljmu.ac.uk/~igm/BAHAMAS/}} \citep{McCarthy2017,McCarthy2018}, EAGLE  \footnote{\url{http://icc.dur.ac.uk/Eagle/}} \citep{Schaye2015,Crain2015,McAlpine2016,Hellwing2016,EAGLE2017}, Illustris  \footnote{\url{https://www.illustris-project.org/}} \citep{Vogelsberger2013,Vogelsberger2014,Illustris2014,Sijacki2015}, and Illustris TNG-300 \footnote{\url{https://www.tng-project.org}} \citep{Springel2018,Pillepich2018,Nelson2018,Naiman2018,Marinacci2018,Nelson2019}. In the case of BAHAMAS, we consider two additional AGN feedback calibrations, dubbed as ``low-AGN'' and ``high-AGN'': in the first one the temperature at which the AGN is activated is lower and thus the AGN feedback is weaker; whereas in the latter AGN feedback is stronger with respect to the standard run.

In Fig.~\ref{fig:data} we show the baryonic effects on the power spectrum, bispectrum, and reduced bispectrum. We display the ratio of the clustering measured in the full hydrodynamical simulations to that in their gravity-only counterparts. Different colours show the results for different simulations, as indicated by the legend.

We see that the amplitude of baryonic effects considerably varies among simulations, in both power spectra and bispectra. In particular, Illustris and Bahamas high-AGN show the strongest suppression in both these statistics, likely due to their strong Supernovae and AGN feedback. On the contrary, EAGLE and Illustris TNG-300 show the smallest baryonic effects
also likely related to their comparatively weak feedback in massive halos. We highlight that both of these simulations display an enhancement of the bispectrum at $k \approx 2-3 \, \ihMpc$, which has been linked to the presence of late-time reaccreted gas by \cite{Foreman2019}. We will test this hypothesis with our baryonification framework later on.

Interestingly, whereas baryons can either suppress or enhance the gravity-only bispectrum, they appear simpler in the reduced bispectrum: baryons always enhance $Q(k)$ on small scales. Qualitatively, there seems to be a clear correlation between the baryonic effects in the power spectrum and bispectrum. However, this correlation is not perfect: Illustris and BAHAMAS high-AGN show similar effects on the reduced bispectrum, but the effects on the power spectrum are clearly different \footnote{However, note that Illustris simulates a box less than $75 \,\hMpc$, thus their results could be affected by cosmic variance and lack of long wavemodes. As showed in Appendix \ref{app:convergence}, massive haloes contribute to baryonic effects more in the bispectrum than in the power spectrum. As a consequence, the reduced bispectrum measured in relatively small boxes is suppressed at small scales with respect to larger boxes.}. In the next sections we will explore whether baryonification methods can successfully describe all these features at high precision.

\begin{figure}
  \includegraphics[width=\linewidth]{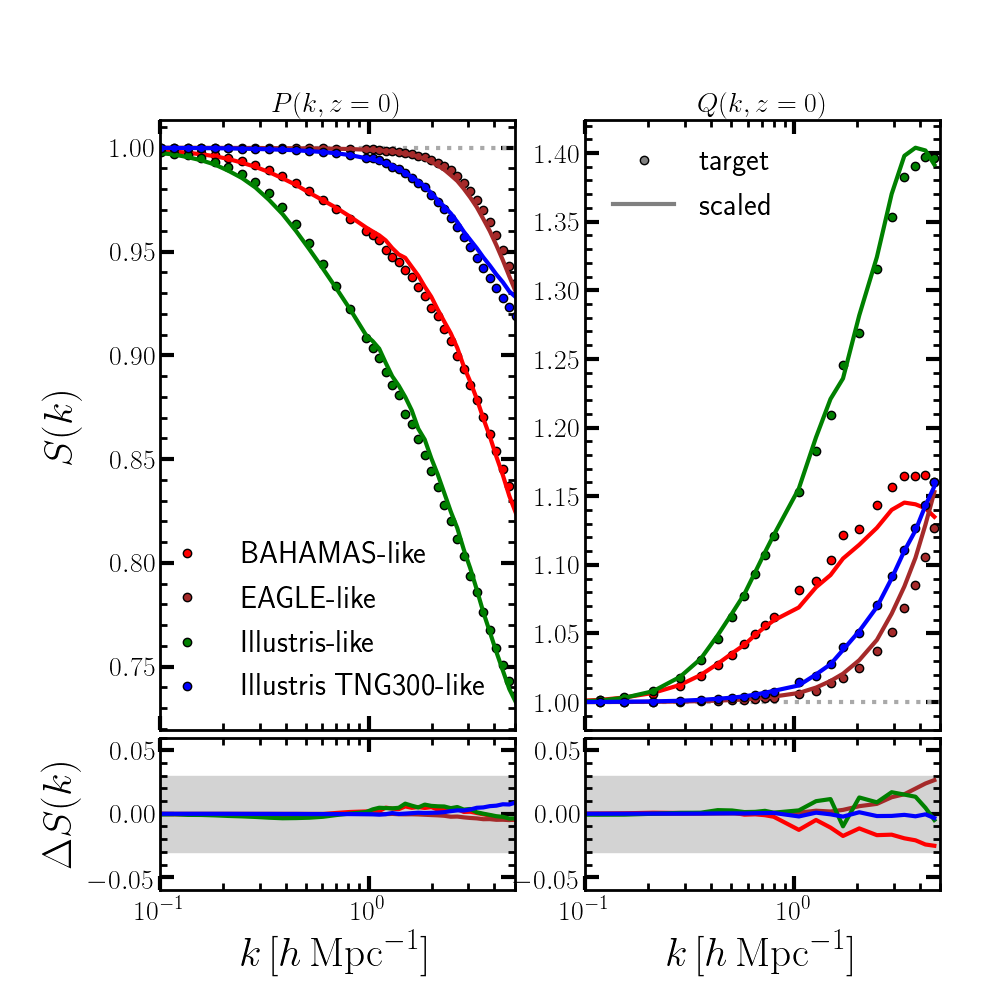}
  \caption{ Accuracy of the cosmology-rescaling algorithm when used with a baryonification procedure. [{\it Upper panel:}] Ratio of the baryonified and gravity-only mass power spectra and reduced bispectra at $z=0$. Symbols show the results using a simulation adopting the {\it Planck} cosmology, whereas lines indicate the results using a simulation {\it rescaled} to the same {\it Planck} cosmology. We provide results for 4 baryonification models roughly consistent with the effects expected in BAHAMAS (red), EAGLE (brown), Illustris (green) and Illustris TNG-300 (blue) models. [{\it Lower panel:}] Difference between the baryonic effects measured in the target and scaled simulation shown in the upper panels. The grey shaded band marks a discrepancy of 3\%.}
  \label{fig:scaling_z0}
\end{figure}

\section{Modelling of the density field}
\label{sec:mbcm}

Given a particle field in a gravity-only (GrO) N-body simulation, we can obtain the
mass field in arbitrary cosmologies and baryonic scenarios by manipulating the
positions and masses of the particles. To do so, we use the framework described in A20, which we recap next.

We first apply a ``cosmology rescaling'' to obtain a simulation at a desired cosmological parameter set \citep{A&W2010}. For this, we scale the lengths, masses and time (by selecting different snapshots) of our simulation in order to match the amplitude of the linear density fluctuation of another cosmology. This technique has been extensively tested \citep[][]{Ruiz2011,Renneby2018,Mead2014a,Mead2014b,Mead2015,AnguloHilbert2015,Zennaro2019,Contreras2020}
providing a $<3\%$ accuracy in the matter power spectrum and $5\%$ in the matter bispectrum up to $k \sim 5\hMpc$ \citep[][Zennaro et al. in prep]{Contreras2020}, over a broad range of cosmologies, even beyond-$\Lambda$CDM. Note we expect a higher accuracy for the ratio of baryonified over gravity-only outputs, as we will show later.

We then apply a ``baryonification'' algorithm to further displace the particles of the simulation, and mimic the effect of different baryonic components. In A20, each halo was assumed to have four components: dark matter, a central galaxy, bound gas, and expelled gas.
In this work, we additionally model satellite galaxies and late-time reaccreted gas, which we describe in detail in the next subsection. The density profiles of all these components are parametrised with physically motivated functional forms, whereas the GrO halo density profile is modelled with as a NFW profile \citep{NFW1997}. Once we have the initial and the ``baryonic'' density profiles, we compute a displacement field which, applied to the halo particles, distorts their distributions accordingly.

\begin{figure}
  \includegraphics[width=0.99\linewidth]{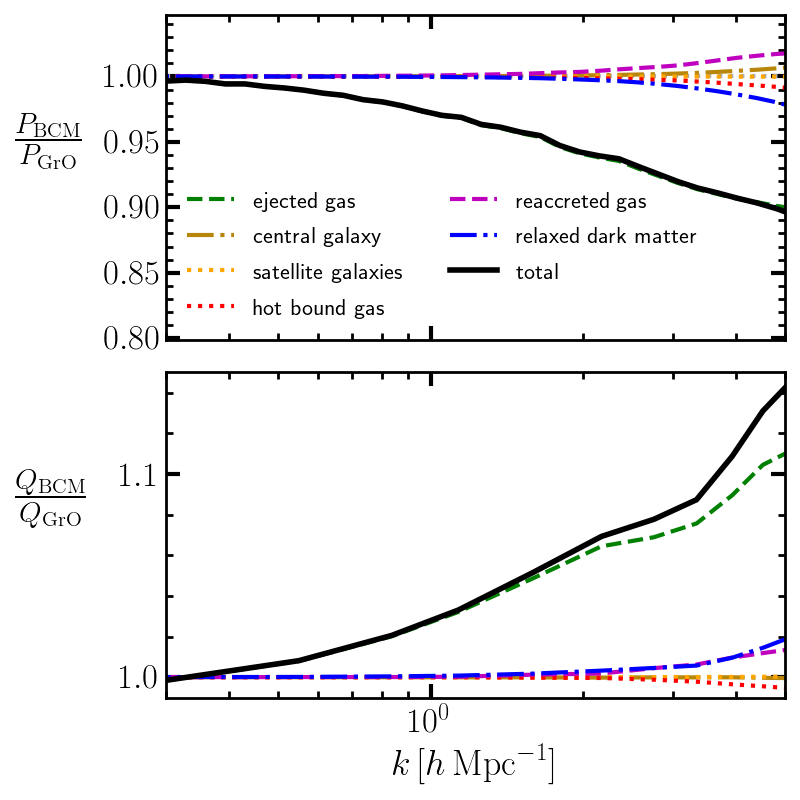}
  \caption{Modifications to the matter power spectrum (upper panels), and reduced bispectrum (lower panels) at $z=0$ caused by baryons, as predicted by our baryonification algorithm with parameters mimicking the effects expected in the Illustris TNG-300 simulations. The total baryonic effect is decomposed into the contribution of each component, namely ejected gas, galaxies, hot bound gas, reaccreted gas, and dark matter, according to the legend.}
  \label{fig:single_component}
\end{figure}

\subsection{Updates of the baryon correction model}
\label{subsec:updates}

One of the main advantages of the {\it baryon correction model} is its extreme flexibility, which allows us to make modifications or include new physics according to various possible scenarios. In this work, we have implemented in the model of A20 the following four main updates:

\begin{itemize}
\item{The adoption of a more flexible functional form for the bound gas;}
\item{The inner slope of the power-law in the central galaxy is a new free parameter;}
\item{The modelling of a satellite galaxies component;}
\item{The inclusion of a late-time reaccreted gas component;}
\end{itemize}

We find that the parametrisation of the bound gas density shape used in \cite{Schneider&Teyssier2015,Arico2020} is not flexible enough to match the profiles measured in a wide range of halo masses of hydrodynamical simulations. We therefore use here a more flexible shape, with an explicit dependence on the halo mass. The shape of the bound gas now reads:

\begin{equation}
\rho_{\rm BG}(r) = \frac{y_0}{ (1+r/r_{\rm inn})^{\beta_{i}} } \frac{1}{(1+(r/r_{\rm out})^2)^2}
\label{eq:rho_bg}
\end{equation}

\noindent where $y_0$ is a normalisation factor, obtained by imposing $\int_{0}^{r_{200}}  {\rm d}r 4 \pi r^2 \rho_{\rm BG}(r) = f_{\rm BG}M_{200}$. The profile is a double power-law with two characteristic scales, $r_{\rm inn}$ and $r_{\rm out}$, defining where the slope changes at small and large radii, respectively. We define the inner radius $r_{\rm inn}=\theta_{\rm inn} \times r_{200}$ and  $r_{\rm out}=\theta_{\rm out} \times r_{200}$, with $\theta_{\rm inn}$ and $\theta_{\rm out}$ being free parameters of the model. The gas inner slope explicitly depends on halo mass as $\beta_{i}=3-(M_{\rm inn}/M_{200})^{\mu_{i}}$, with the characteristic mass $M_{\rm inn}$ and $\mu_{i}$ as free parameters. After checking the small impact that $\mu_{i}$ has on both power spectrum and bispectrum,
 we have fixed its value to $\mu_{i}=0.31$, in agreement to the Model A-avrg in \cite{Schneider&Teyssier2019}.

This profile is similar to that in \cite{Schneider&Teyssier2019}, with the main difference being  that in our model the bound gas perfectly traces the dark matter on scales beyond $r_{\rm out}$ and the ejected gas decay exponentially, whereas \cite{Schneider&Teyssier2019} models a single gas component, with the slope at large radii as a free parameter.

The central galaxy density profile is given by

\begin{equation}
\rho_{\rm CG}(r) = \frac{y_0}{ R_h  r^{ \alpha_g}} \, \exp \left[- \left( \frac{r}{2R_h} \right)^2 \right],
\label{eq:rho_cgalaxy}
\end{equation}

\noindent where $y_0$ is found imposing $\int_{0}^{r_{200}} \rm{d^3} \rho_{\rm CG}(r) = f_{\rm CG}M_{200}$, the half-mass radius is $R_h = 0.015 \times r_{200}$ and $\alpha_g$,
the inner slope of the central galaxy, is a free parameter of the model, with a fiducial value $\alpha_g=2$.

In addition to the central galaxy, we add the stellar component of satellite galaxies. Stellar mass is, to a good approximation, collisionless and thus a good tracer of dark matter. For this reason we model the contribution of satellite galaxies as the dark matter. The dark matter back-reacts to the baryonic potential well, and therefore the satellite galaxies, being a linearly biased tracer of the dark matter, are quasi-adiabatically relaxed. We refer the reader to A20 for the details of the implementation of the back-reaction mechanism.

\begin{figure}
  \includegraphics[width=.99\linewidth]{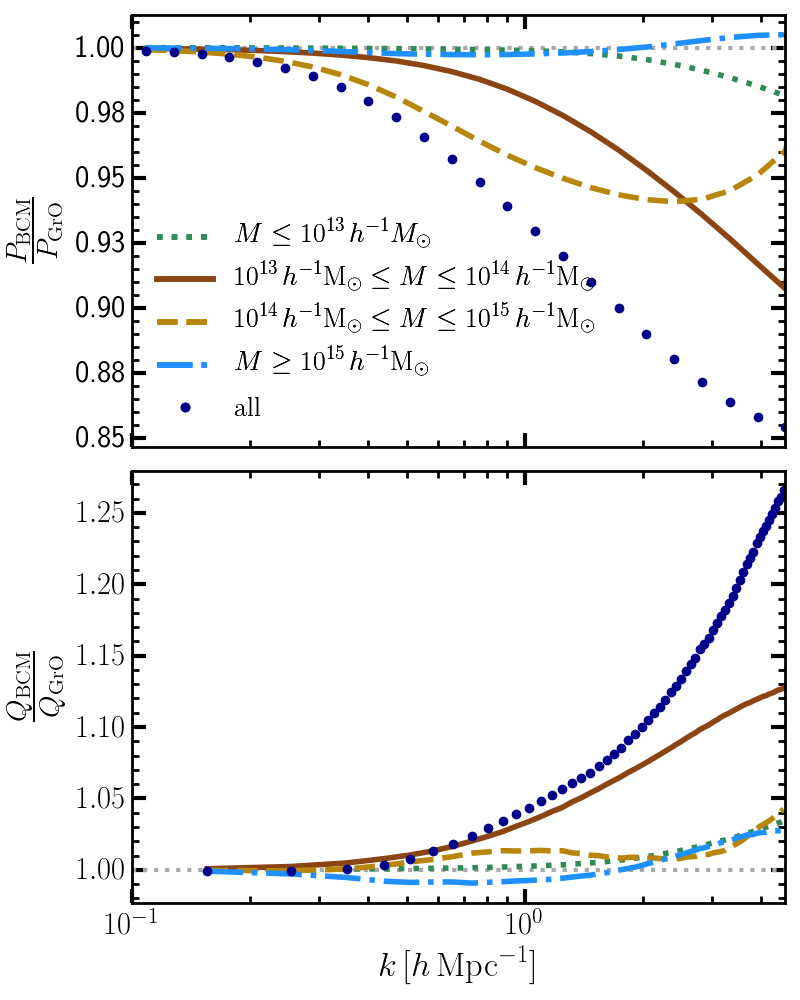}
  \caption{ {\it Upper panel:} Baryon suppression of the matter power spectrum at $z=0$, when applying the baryon correction model to haloes smaller than $10^{13} \Msun$ (green dotted line), between $10^{13}-10^{14}\Msun$ (brown solid line), $10^{14}-10^{15}\Msun$ (golden dashed line) and to all the haloes (blue dots) of our $512 \, \hMpc$ simulation.
  {\it Lower panel:} Same as the upper panel, but for the ratios between baryonic and gravity-only results in the reduced bispectrum.}
\label{fig:mass_bins}
\end{figure}

Motivated by the hypothesis of \cite{Foreman2019}, who suggested the presence of an overdensity of gas reaccreted at late times to explain the maximum in the bispectrum around $k \approx 2.5 \ihMpc$ in the Illustris TNG-300 simulation, we added to our model a new gas component, which mimics such gas overdensity. We assume this new component to be Gaussian shaped:

\begin{equation}
\rho_{\rm RG}(r) = \frac{y_0}{\sqrt{2 \pi} \sigma_r} \exp \left[ - \frac{(r - \mu_r)^2}{ (2 \sigma_r)^2} \right],
\end{equation}

\noindent where $y_0 = f_{\rm RG} M_{200} /  \int_{0}^{r_{200}} 4 \pi r^2 \rho_{\rm RG}(r)  dr $.

For simplicity, we assume the gas overdensity to have a fixed spatial distribution in terms of the halo virial radius, $\mu_r= 0.3 \times r_{200}$ and  $\sigma_r= 0.1 \times r_{200}$, and after checking that our main results are not affected by this choice. We let free
instead the mass fraction, $f_{\rm RG}$, as explained in what follows.

All the density profiles of the baryon correction model are normalised to $M_{200}$, with the abundance of each component determined by its respective mass fraction. The dark matter fraction is fixed by cosmology, $f_{\rm DM} = 1-\Omega_{\rm b}/\Omega_{\rm m}$.

The central galaxy fraction is given by an abundance-matching parametrisation \citep{Behroozi2013}:

\begin{equation}
 f_{\rm CG} (M_{200}) = \epsilon \left( \frac{M_1}{M_{200}} \right) 10^{g(\log_{10}(M_{200}/M_1)) - g(0)},
\label{eq:f_cg}
\end{equation}

 \begin{equation}
    g(x)= -\log_{10} (10^{\alpha x} +1) + \delta \frac{ (\log_{10} (1+\exp(x)))^\gamma}{1+\exp(10^{-x})}.
 \end{equation}

We use the best-fitting parameters at $z=0$ given by \cite{Kravtsov2018}, along with the redshift dependence given by \cite{Behroozi2013}, both reported in Appendix A of A20 and not included here for the sake of brevity.

Satellite and central mass fractions have the same parametric form, and their parameters are
assumed to be linearly dependent e.g. $M_{\rm 1,sat} (z=0)= \alpha_{\rm sat} M_{\rm 1,cen}(z=0)$, with $\alpha_{\rm sat}$ as a free parameter of the model, similar to the approach of \cite{Watson2013}.

The halo gas mass fraction, defined as the sum of the bound gas and the reaccreted gas, is

\begin{equation}
f_{\rm HG} (M_{200}) = f_{\rm BG}+f_{\rm RG}= \frac{\Omega_b/\Omega_m-f_{\rm CG}-f_{\rm SG}}{1+(M_c/M_{200})^\beta},
\label{eq:f_hg}
\end{equation}

\noindent with $M_c$ and $\beta$ free parameters, and $f_{\rm CG}$, $f_{\rm SG}$, the central and satellite galaxy mass fractions, respectively.
The reaccreted gas mass fraction is

\begin{multline}
f_{\rm RG} (M_{200}) = \frac{\Omega_b/\Omega_m-f_{\rm CG}-f_{\rm SG}-f_{\rm HG}}{1+(M_r/M_{200})^{\beta_r}} = \\ = f_{\rm HG} \frac{(M_c/M_{200})^{\beta}}{1+(M_r/M_{200})^{\beta_r}} ,
\label{eq:f_rg}
\end{multline}

\noindent with $M_r$ as a free parameter and $\beta_r$ fixed for simplicity to $\beta_r=2$.

Finally, the bound and ejected gas mass fractions are set by mass conservation:

\begin{equation}
f_{\rm BG} = f_{\rm HG}-f_{\rm RG};
\label{eq:f_bg}
\end{equation}
\begin{equation}
f_{\rm EG} = \Omega_b/\Omega_m-f_{\rm CG}-f_{\rm SG}-f_{\rm HG}.
\label{eq:f_eg}
\end{equation}

As an example, we show in Fig.~\ref{fig:density_profiles} how this updated model is able
to reproduce at the same time the gas and stellar density profiles
measured in three different halo mass bins of Illustris TNG-300, $[10^{13.5}-10^{14}] \, \Msun$,
 $[10^{14},10^{14.5}] \, \Msun$ and  $[10^{14.5},10^{15}] \, \Msun$. Note that here our reaccreted mass fractions are consistent with zero, thus not appearing in the plot.

\begin{figure*}
  \includegraphics[width=0.99\linewidth]{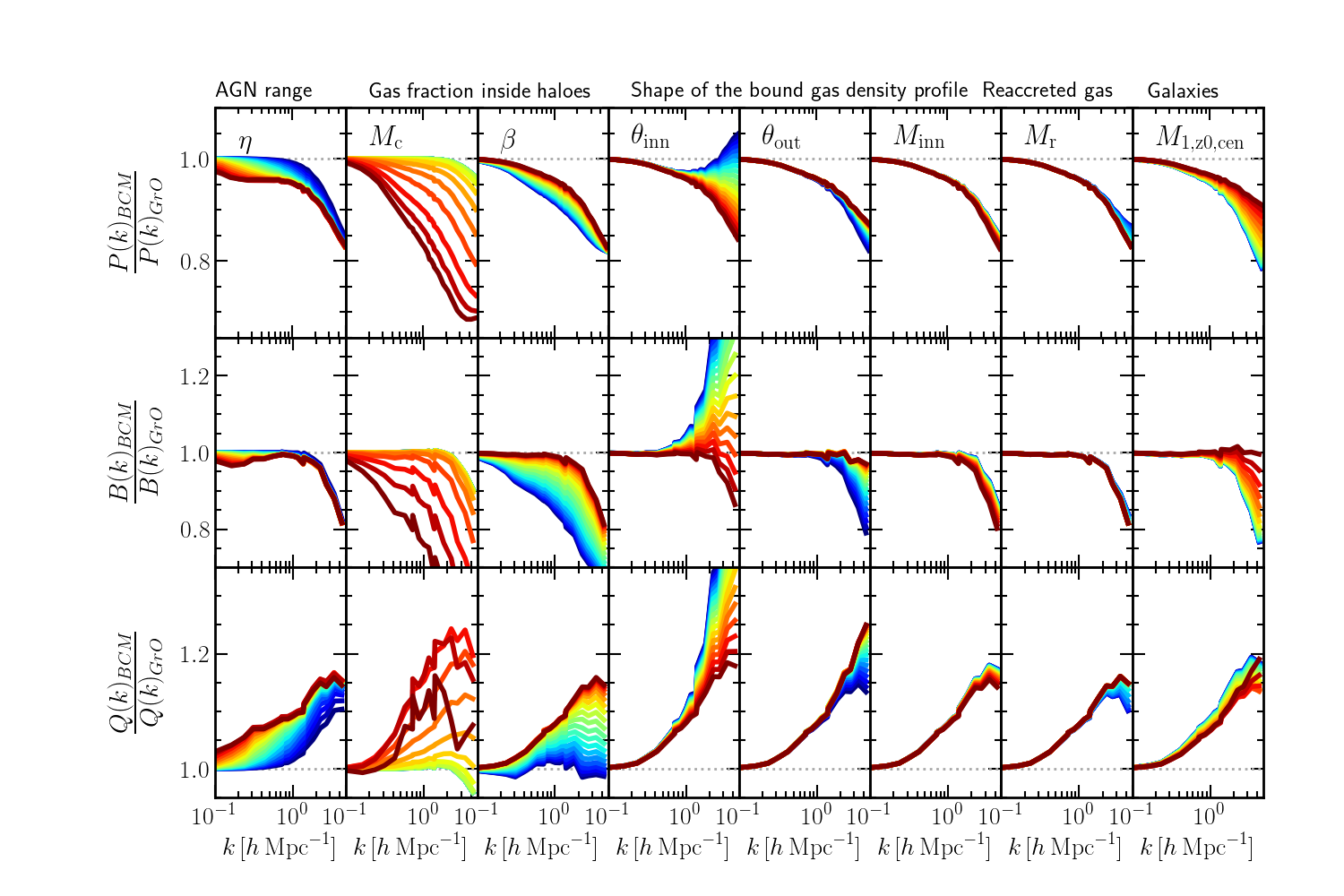}
  \caption{Modifications to the matter power spectrum (upper panels), bispectrum (central panels) and reduced bispectrum (lower panels) at $z=0$ caused by baryons, according to our baryonification algorithm. Each column varies one of the free parameters of the model while keeping the others fixed at their fiducial value. Parameter ranges are $\log M_c\in[9,15]$,
$\log \eta\in[-0.7,0.7]$, $\log \beta\in[-1,0.7]$, $\log M_{\rm r}\in[12,17]$, $\log  M_{\rm 1,z0,cen}\in[9,13]$, $\log \theta_{\rm inn}\in[-2,-0.5]$, $\log  \theta_{\rm out}\in[-0.5,0]$, $\log  M_{\rm inn}\in[12,16]$.
Blue to red colors denote low to high parameter values.}
  \label{fig:bcm_dependence}
\end{figure*}
\subsection{Accuracy of cosmology rescaling and baryonification in the bispectrum}
\label{subsec:cosmoscaling}

In A20 we showed that applying a baryonification algorithm together with a cosmology-rescaled simulation leaded to percent-accurate results in the power spectrum. We now perform an analogous test to validate the performance of the updated model and extend the analysis to the bispectrum.

In Fig.~\ref{fig:scaling_z0} we compare the baryonic effects on the power spectrum and reduced bispectrum as measured in a simulation carried out with a {\it Planck} cosmology and a simulation carried our with a {\it Nenya} cosmology and then rescaled to a {\it Planck} cosmology (c.f. \S\ref{sec:sim}). These two cases are denoted as target and scaled, respectively, and displayed by symbols and lines as indicated by the legend.

We display 4 different baryonification parameter sets, chosen to roughly reproduce the clustering of EAGLE, Illustris, Illustris TNG-300 and BAHAMAS. We can see that the difference between applying the BCM on top a rescaled or target simulation is less than $1\%$ in the power spectrum and less than $3\%$ in the reduced bispectrum. We show these results only for $z=0$, but we have explicitly checked that at higher redshifts we obtain similar outcomes.

We note that the initial conditions of the target simulation were not set to match that of the simulation we scale, nor its volume have been chosen to match the volume of the rescaled simulation (which could have increase the agreement further). Nevertheless, the errors
we obtain are comparable to our target accuracies for reproducing the baryonic effects on the power spectrum and bispectrum.

\section{Impact of baryons on the bispectrum}
\label{sec:impact}

In this section we systematically explore the effects that the various baryonic components, and the free parameters associated to them, produce on the clustering.

We first isolate the effect of each baryonic component by selecting them one-by-one and considering all the others collisionless (thus behaving like dark matter). As shown in Fig.~\ref{fig:single_component}, we find that, in agreement with \cite{Schneider&Teyssier2015,Arico2020}, the ejected gas largely dominates the suppression in the power spectrum, despite its low mass fraction. Interestingly, the ejected gas shows the largest effect also in the reduced bispectrum, but in this case it causes an enhancement of the power at all scales.

As an qualitative explanation, let us consider two overdensity fields, $\delta_{\rm BCM}$ and $\delta_{\rm GrO}$. Assuming that one is suppressed with respect to the other,
$\delta_{\rm BCM}= (1-\alpha)\delta_{\rm GrO}$, it is easy to show that the ratios between their power spectra and equilateral bispectra are $P_{\rm BCM}/P_{\rm GrO} = (1-\alpha)^2$, and $B_{\rm BCM}/B_{\rm GrO} = (1-\alpha)^3$, respectively. Therefore, the reduced bispectrum ratio is $Q_{\rm BCM}/Q_{\rm GrO} = (1-\alpha)^{-1}$.
In other words, we observe an enhancement of the reduced bispectrum because the suppression in the bispectrum is smaller than the squared suppression of the power spectrum. The other components are, in this particular setting of the BCM which roughly mimics the BAHAMAS simulation, subdominant, contributing to about $2\%$ in the power spectrum and reduced bispectrum. The reaccreted gas, in particular, causes an enhancement at small scales in both the power spectrum and reduced bispectrum.

It is interesting to explore which halo masses contribute the most to the baryonic effects on clustering. In order to answer this question, we have split the halo catalogue of our simulation in different mass bins, and then we have applied our BCM separately to each of them. In Fig.~\ref{fig:mass_bins} we show how haloes between $10^{13}-10^{14}\Msun$ contribute  more than $50\%$ of the effect on the power spectrum at small scales. Haloes of $10^{14}-10^{15}\Msun$ are dominant at large scales in the power spectrum, whereas at the small scales, slightly smaller haloes contribute more. Haloes with $M<10^{13}\Msun$ and $M>10^{15}\Msun$ contribute for less than $2\%$ percent, and only at small scales.

The relative contribution of halos of different mass slightly changes in the case of the reduced bispectrum. The dominant contribution is still from haloes of $10^{13}-10^{14}\Msun$, but the relative impact of the most massive haloes in the simulation ($M>10^{15}\Msun$) is not as small as for the power spectrum. The fact that the bispectrum is more sensitive to the largest haloes is not surprising \citep[see e.g.][]{Foreman2019}, and has as a practical outcome the slower convergence of the bispectrum with simulated volume compared to that of power spectrum, which we investigate in Appendix \ref{app:convergence}.

We quantify now the impact of the free parameters in the power spectrum and reduced bispectrum. To do so, we vary each parameter one by one, while keeping the others
fixed to the value that best fits the BAHAMAS (standard AGN) simulation, described in \S\ref{sec:hydro_bcm}. The intervals in which we vary parameters, in $log_{10}$, are the following: $M_c\in[9,15]$, $\eta\in[-0.7,0.7]$, $\beta\in[-1,0.7]$, $M_r\in[12,17]$, $M_{\rm 1,z0,cen}\in[9,13]$, $\theta_{\rm inn}\in[-2,-0.5]$, $\theta_{\rm out}\in[-0.5,0]$, $M_{\rm inn}\in[12,16]$.

Note we do not show any free parameters for satellite galaxies, as they have a negligible impact on the matter clustering. In fact, they are a biased tracer of the dark matter, and additionally their mass fraction is very small. Given that the relaxation of the dark matter contributes only for a few percent in the matter clustering, it is easy to see why the baryonic effect caused by satellite galaxies is negligible. Thus, we fix their values to the best-fitting of the stellar profile of the Illustris TNG-300 simulation found in \S\ref{subsec:updates}.

In Fig.~\ref{fig:bcm_dependence} we display the mass power spectra, bispectra, and reduced bispectra obtained after applying the BCM to a GrO $N$-body simulation. Each panel varies a single parameter of the model while keeping the others fixed to their fiducial value. Bluer (redder) colors represent low (high) parameter values. We can see that almost all the parameter combinations predict a suppression in the power spectrum and an enhancement on the reduced bispectrum, at all the scales. Specifically, by increasing $\eta$ (the parameter which set the maximum range of the AGN feedback), the suppression (enhancement) of the
power spectrum (reduced bispectrum) is pushed, as expected, towards larger scales.
The parameters $M_c$ and $\beta$ set the fraction of gas which is retained in haloes of a given mass, and thus also the mass of gas that is expelled. Therefore, is not surprising that these parameters have a big impact on both power spectrum and bispectrum, given that the ejected gas component is the dominant one. Varying $M_c$ we span a $30\%$ range in the clustering; in particular, higher values mean that increasingly larger haloes are free of gas, thus more ejected gas. In these cases we see, accordingly, a larger suppression in the power spectrum and enhancement in the reduced bispectrum.

Varying the shape of the bound gas through the parameters $\theta_{\rm inn}$, $\theta_{\rm out}$ and $M_{\rm inn}$ has an impact only on small scales. Specifically, the model seems very sensitive to $\theta_{\rm inn}$, for which we see a substantial enhancement of both power spectrum and reduced bispectrum when changing the inner gas slope at increasingly smaller radii. On the other hand, the dependence on $M_{\rm inn}$ looks negligible. As expected, increasing  $M_{\rm 1,z0,cen}$, and thus having the peak of the star formation at higher halo masses, results in more power at small scales.

Finally, we see that the impact of the late-time reaccreted gas is very modest, despite we vary its mass fraction from practically zero to a limit value of $\approx50\%$ for some halo masses. Arguably, the effect of this parameter in both power spectrum and reduced bispectrum, can be absorbed by a combination of the other parameters of the model, but might become more important on smaller scales.

We note that the models shown in  Fig.~\ref{fig:bcm_dependence} are just an illustrative example, and do not encompass all the possible modifications given by the BCM: even if the described trends would be likely similar, changing the underlying fiducial model would result in different amplitude and shapes of baryonic effects.

\begin{figure*}
\centering
  \includegraphics[width=0.99\linewidth]{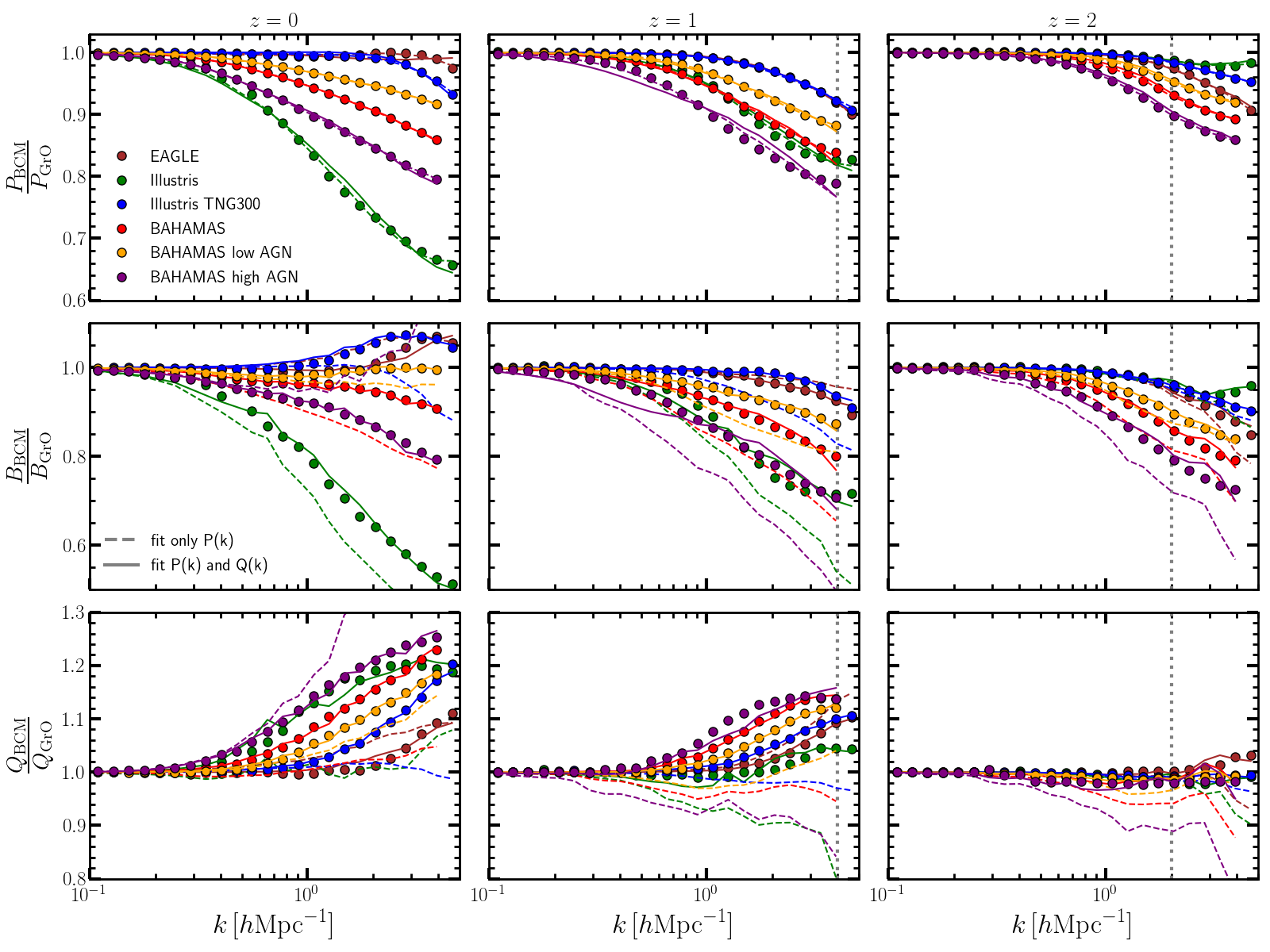}
  \caption{Measurements of the baryonic impact to the matter power spectrum, $S(k) \equiv P/P_{\rm GrO}$ (upper panels), equilateral bispectrum, $S(k) \equiv B/B_{\rm GrO}$ (central panels), and reduced equilateral bispectrum, $S(k) \equiv Q/Q_{\rm GrO}$ (bottom panels), in different hydrodynamical simulations according to the legend (symbols), at $z=0$ (left), $z=1$ (centre) and $z=2$ (right). The best-fitting baryonification model constrained using only the power spectrum is displayed as dashed lines, whereas the best-fitting model constrained on both the power spectrum and reduced bispectrum shown with solid lines. Grey vertical dotted lines mark the scales where the estimated shotnoise contributes to $>1/3$ of the clustering amplitude.}
  \label{fig:hydro_models}
\end{figure*}

\begin{figure*}
  \includegraphics[width=.85\linewidth]{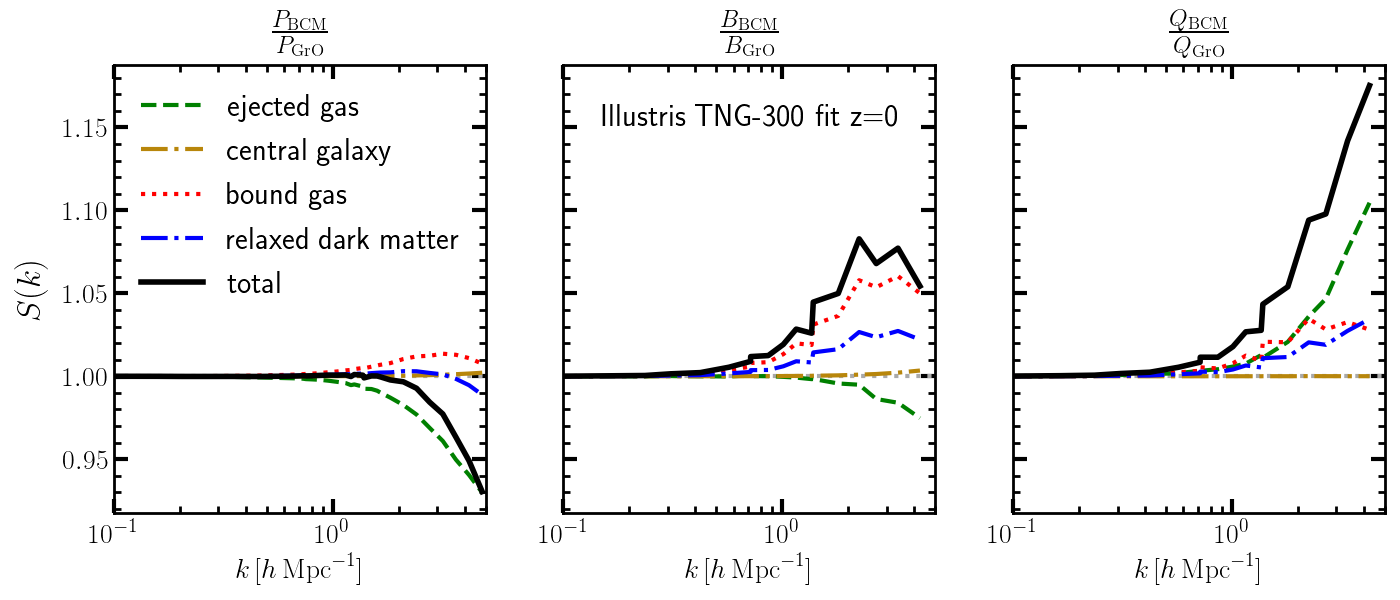}
  \caption{Best-fitting model to Illustris TNG-300 power spectrum and bispectrum, at $z=0$. The contribution of the varyous baryon component are isolated in the power spectrum (left), equilaetral bispectrum (centre) and reduced bispectrum (right). Note that the bound gas, and the back-reaction to the dark matter, are the principale causes of the bump visible in the bispectrum at small scales.}
          \label{fig:tng_components}
\end{figure*}

\begin{figure}
  \includegraphics[width=.99\linewidth]{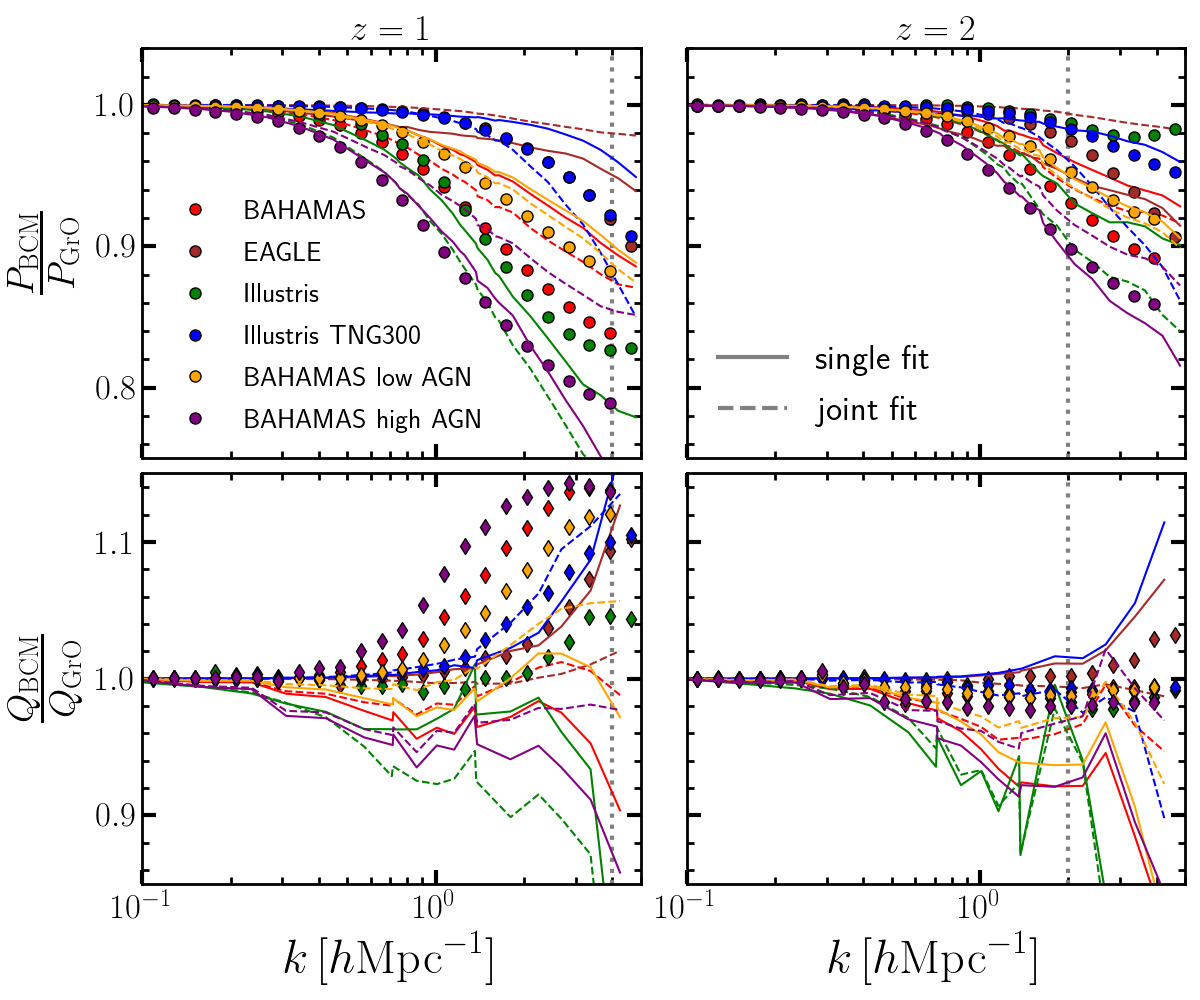}
  \caption{Impact of the redshifts evolution in the baryonic correction parameters.
          We have fitted the power spectra (upper panels) and reduced bispectra (lower panels) of the hydrodynamical simulations reported in the legend at $z=0$, and then applied the same model at higher redshifts $z=1,2$, assuming our best-fitting parameters to be redshift independent.}
          \label{fig:fixed_z0}
\end{figure}

\section{Fitting the hydrodynamical simulations}
\label{sec:hydro_bcm}

In this section, we explore whether the BCM is able to reproduce the impact of baryons in  six different state-of-the-art hydrodynamical simulations, namely EAGLE, Illustris, Illustris TNG-300, and three different AGN implementations of BAHAMAS. We remind the reader that these simulations differ in cosmology, $N$-body code, sub-grid physics, box size, and observables with which they have been calibrated. They show a difference of $30\%$ at $z=0$ in the power spectrum and $15\%$ in the reduced bispectrum, thus being a good benchmark for the flexibility and realism of our model.

We fit the power spectrum and the reduced bispectrum of each hydrodynamical simulation over the range $0.1 < k/ (\ihMpc) < 5$,  both separately and jointly, varying seven free parameters: $M_c$, $\eta$, $\beta$, $M_{\rm 1,z0,cen}$, $\theta_{\rm inn}$, $\theta_{\rm out}$, $M_{\rm inn}$ within the priors shown in \S\ref{sec:impact}. We assume no correlation among power spectrum and bispectrum nor among the measurements at different wavenumbers.
Specifically, we use an empirical approach similarly to A20, where the covariance matrix is directly estimated by the intra-data variance, giving the same weights to power spectrum and bispectrum.
We expect the errors associated to the bispectrum ratios to be larger than the one
associated to the power spectrum (see for instance the errors measured by \cite{Foreman2019} by dividing the hydrodynamical simulation volume in subboxes). Nevertheless, being the purpose of this test to asses
the accuracy of the joint fit of power spectrum and bispectrum, we avoid to give more weight to the former to not
degrade the fit of the latter.

To perform the fit, we have implemented a {\it particle swarm optimisation} algorithm \citep{Kennedy1995}. In this algorithm, a pack of particles efficiently searches the minimum of a function in a given parameter space. Each particle communicates with the others at each step, and they are attracted both to their local and the swarm global minima, with a relative strength that can be tuned. The velocity and position of the particles are updated in every step, depending solely on the swarm status in the previous step. For our application, we use a swarm of $10$ particles and $250$ iterations, finding that an average of 100-150 steps are enough to converge to the global minimum.

In Fig.~\ref{fig:hydro_models} we present the main result of this paper. We show the best BCM fits at three different redshifts, $z=0$, $1$,  and $2$. We have marked with a grey dotted line the scales where we estimate the shotnoise amplitude to be approximately 30\% of the clustering amplitude: $k\approx 2\ihMpc$ at $z=2$, and $k\approx 4\ihMpc$ at $z=1$. We remind the reader that, in the cosmology rescaling process, the box of the simulations can vary of length, and so the shotnoise level can be slightly different. Due to the significant contribution of shotnoise, results at small scales and high redshifts should be interpreted carefully.

Dashed lines show the results when fitting only the power spectrum measurements.
In this case, we recover the accuracy of $1\%$ found in A20 in the power spectrum at all scales and redshifts. However, the baryonic impact on the bispectrum can be over- or under-estimated by up to 20\%. In contrast, when fitting the power spectrum and bispectrum {\it together}, the accuracy of the power spectrum slightly degrades, but we obtain significantly better agreement with the bispectrum.

For all redshifts and hydrodynamical simulations considered, we obtain joint fits that are $1-2\%$ accurate for the power spectrum and $3\%$ for the bispectrum. We note that the worse performance is obtained for the simulations with the most extreme feedback e.g. Illustris and BAHAMAS high-AGN. In the case of BAHAMAS, which is arguably the most realistic simulation for our purposes, the fits describe simultaneously the baryonic effects on the power spectrum and bispectrum to better than $1\%$.

These results are achieved considering the late-time reaccreted gas fixed to zero.
In particular, we note that the bump around $k\approx2-3 \ihMpc$ in the bispectrum of
Illustris TNG-300 and EAGLE are correctly reproduced by the model, despite the absence of the reaccreted gas component. To understand which BCM component causes this enhancement of
the bispectrum at small scales, we isolate the impact of each baryon component
to the clustering, similarly to what done in \S~\ref{sec:impact}. This analysis, reported in Fig.~\ref{fig:tng_components}, clearly show that the bound gas causes an enhancement at small scales in both power spectrum and bispectrum. Furthermore, the back-reaction of the gas overdensity to the dark matter adds the necessary power to reproduce correctly the measurements of the Illustris TNG-300. We can conclude that, by simply assuming the gas as a double power-law, the model has enough flexibility (over the range of scales we considered) to explain the ``bump'' in the bispectrum measured in Illustris TNG-300 and EAGLE.

 We have repeated the fits letting free the corresponding mass fraction, $M_r$, however, this did not result in noticeably improved fits. This finding is consistent with the hypothesis that, within the accuracy of our model and simulated data, and over the scales considered, the reaccreted gas is not necessary to reproduce the clustering of the hydrodynamical simulations analysed.

Finally, one could wonder what is the smallest number of free parameters necessary to produce accurate results. In A20 it is shown that with only 4 parameters it is possible to fit the power spectrum at $1\%$, and arguably the 7-8 parameters used here are degenerate, and effectively recastable into a model with a smaller parameter set. We leave the exploration of degeneracies between parameters and the finding of a minimal-model parameter set for a future work.

\begin{figure}
  \includegraphics[width=.9\linewidth]{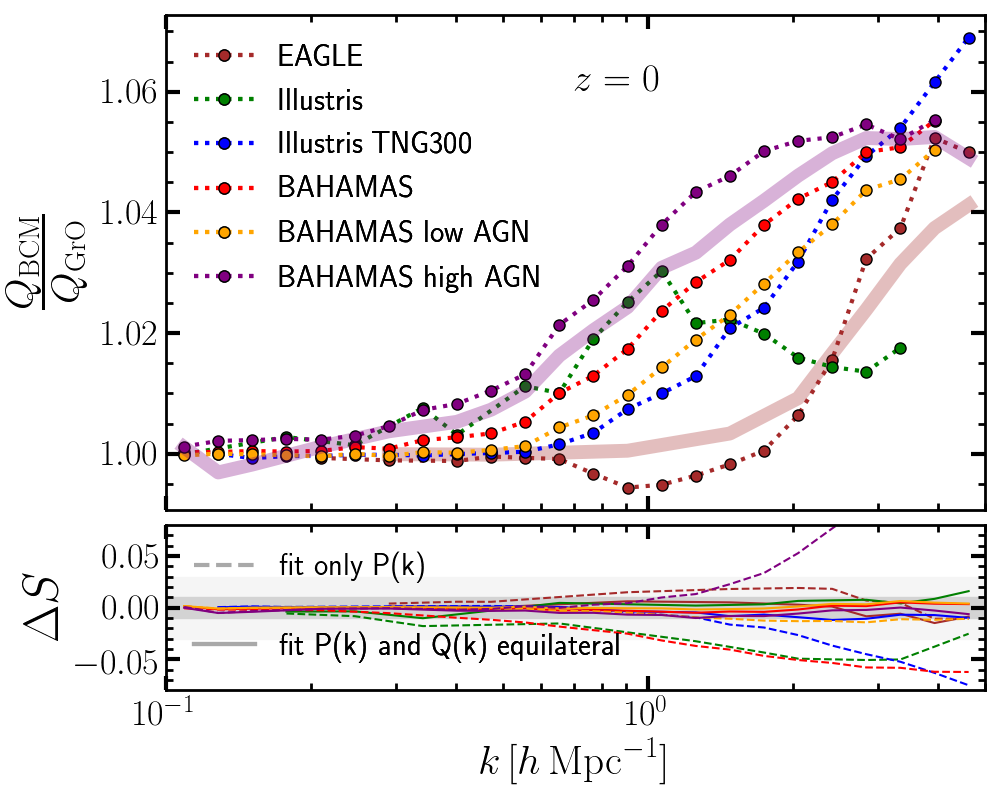}
  \caption{{\it Upper panel:} Measurements of the reduced squeezed bispectra at $z=0$ in EAGLE, Illustris, Illustris TNG-300, BAHAMAS, Bahamas low-AGN and BAHAMAS high-AGN (symbols), according to the legend. The shaded band show our prediction obtained by fitting power spectrum and reduced equilateral bispectrum inthe two most extreme models, EAGLE and BAHAMAS high-AGN. {\it Lower panel:} Difference between the ratios of reduced squeezed bispectra predicted by our baryon correction model and measured in the hydrodynamical simulations. The dashed lines show the model parameters fitting only the power spectra of the hydrodynamical simulations, whereas the solid lines show the model with parameters constrained by fitting both power spectra and equilateral reduced bispectrum.}
  \label{fig:squeezed}
\end{figure}

\subsection{Redshift dependence of the baryon parameters}
\label{subsec:z_dependence}
As already pointed out in previous works \citep{Chisari2019,Arico2020}, despite the baryonic parameters do not have a specific redshift dependence, when fitting the clustering at different redshifts they show a clear evolution. To quantify the inaccuracies obtained by fixing the baryonic parameters, we apply the baryonification algorithm to our simulation to snapshots that correspond to high redshifts ($z=1,2$), using the best-fitting parameters at $z=0$ found in \S~\ref{sec:hydro_bcm}.

In Fig.~\ref{fig:fixed_z0} we show the power spectra and reduced bispectra obtained. The error in the power spectrum is in most of the cases below $5\%$, for extreme models around 7-10$\%$. On the other hand, the reduced bispectrum can be off of $10-20\%$. These errors must be taken into account when fixing at face value a set of baryonic parameters in multiple redshifts.

 \begin{figure}
   \includegraphics[width=.99\linewidth]{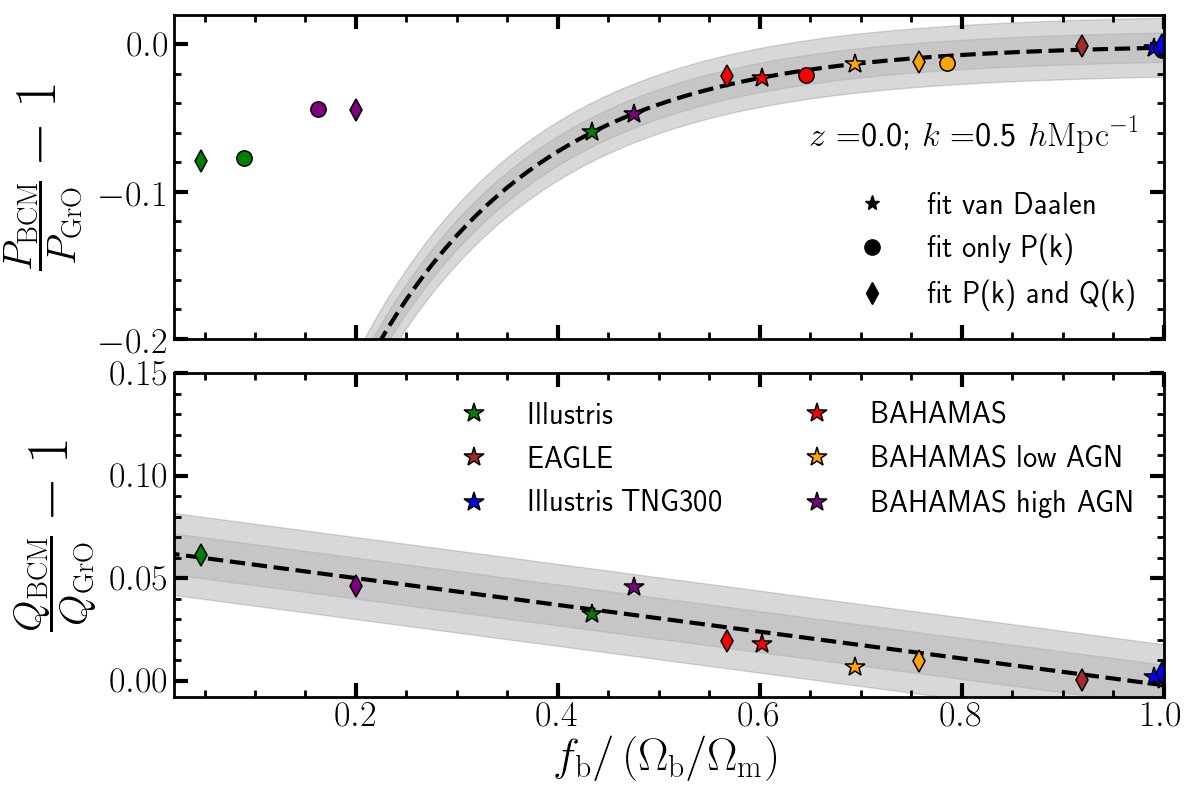}
   \caption{{\it Upper panel}: Relation between the baryonic impact on the power spectrum at $k=1\, \ihMpc$, defined as $\Delta P(k) / P(k)$, and the halo baryon fraction in haloes with a mass [$6\times10^{13}$,$2\times10^{14}$] M$_{\odot}$.
    The black dashed line displays the fit provided by \protect\cite{vandaalen2019}, with the grey and light grey shaded bands marking a 1\% and 2\% deviation, respectively. For comparison, the colored stars emply the baryon fraction measured directly in the hydrodynamical simulations by \protect\cite{vandaalen2019}. The coloured symbols indicate the measurements of our baryonified simulations when using the best-fitting values calibrated against the power spectra (circles), and both power spectra and bispectra (diamonds). {\it Lower panel:} Same as the upper panel, but for the reduced bispectrum. In this case, the dashed line represents a simple linear regression of the symbols displayed.}
   \label{fig:barfraction}
 \end{figure}

\subsection{Baryonic effects on the squeezed bispectrum}

We have so far analysed, for simplicity, only the equilateral configuration of the reduced bispectrum. In this section, we explore the baryonification performance for the ``squeezed'' configuration, which measures the correlation between points on isosceles triangles with one side much smaller than the other two in $k$-space, so that $k_1 \ll k_2 = k_3 $. The squeezed bispectrum might be seen as a ``conditional'' two-point correlation which quantifies the dependence of small-scale nonlinearities on the large-scale background overdensity.

It has been shown that, in some cases, the baryonic effect on the squeezed bispectrum can be directly related to the power spectrum at small scales, when considering a $k_1$ long enough to not be affected by baryonic physics \citep{Barreira2019,Foreman2019}. Specifically, \cite{Barreira2019} have measured the ``power spectrum response functions'' in the Illustris TNG-300, using the separate universe approach, finding that they are largely unaffected by baryonic physics. This suggests that the information in the squeezed bispectrum is already contained in the power spectrum, and thus knowing the latter we can predict the former. However, as shown in \cite{Foreman2019}, the analytical predictions given from the power spectrum response function are not always in agreement with the hydrodynamical simulations, e.g. BAHAMAS. This could be a hint that, in some cases, the response function are not fully specified by the power spectra.

Here, we take a somewhat agnostic approach, and test if we can predict correctly the squeezed bispectrum starting from the information contained in the power spectrum and the equilateral bispectrum. To do so, we apply to our gravity-only simulations a BCM with the parameters that reproduce both the power spectrum and reduced equilateral bispectrum for a given hydrodynamical simulations. Then, we measure the reduced squeezed bispectrum ($k_1 = k_2 > k_1 \sim 0.1\,\hMpc$) and compare it with those measured directly in the hydrodynamical simulations.

In Fig.~\ref{fig:squeezed} we show the results obtained at $z=0$. First, we can notice that, as for the case of equilateral configurations, when considering baryon physics the reduced squeezed bispectrum is enhanced with respect to the gravity-only one. However, the baryonic effects in the squeezed bispectrum are smaller than those in the equilateral configuration -- spanning a range of $2-7\%$, against a $10-25\%$ measured in the equilateral configuration.

We also see that our predictions for the squeezed reduced bispectrum agree very well with the simulation measurements, reaching a $\sim 1\%$ accuracy in all cases. This further supports the idea that the modifications to the density field in the barionification is accurately capturing the three-dimensional distortions induced by baryons, and not simply fitting an effective distortion in the power spectrum.

For comparison, we also display in Fig.~\ref{fig:squeezed} the predictions when tuning our model using only the power spectrum. As for the equilateral bispectrum, the impact of baryons is not captured very accurately, with discrepancies generally within $5\%$ (EAGLE and BAHAMAS low-AGN $\le 2\%$) to up to $10\%$ (Bahamas high-AGN).

\subsection{Baryon fractions in haloes}

Recently, it has been shown that there is a tight correlation between baryonic effects on
the power spectrum and the baryonic fraction inside haloes of $M \approx 10^{14} M_{\odot}$ \citep{vandaalen2019}. The best fits of the baryon correction model has been shown to be able to accurately recover such correlation, even if for large power spectrum suppression, which correspond to very strong AGN feedback, it tends to underestimate the baryon fraction measured in hydrodynamical simulation \citep{Arico2020}.

We now explore whether adding the information on the bispectrum the baryonic halo fractions become more constrained, and additionally, whether there is a relation between reduced bispectrum and baryon fraction, analogous to the one found for the power spectrum.

In Fig.~\ref{fig:barfraction} we show how, indeed, in the case of the BAHAMAS high-AGN simulation, the fit of the bispectrum marginally improves the baryon fraction estimation. The fact that both the best-fitting parameter set can accurately reproduce the power spectra of the hydrodynamical simulations, but predict slightly different baryon fractions, can be a hint of some degeneracies between parameters which is broken when including the bispectrum information. For the Illustris, the opposite is true: the gas fraction in clusters differs more from its true value. This likely points to the fact that some of the baryonification assumptions somewhat break for extreme feedback scenarios. This could be related to gas fractions that are not monotonic with halo mass, or that these events affect gas beyond the boundaries of a halo (a process not included in our model). On the other hand, we note that the gas fractions in these simulations are in clear tension with observations which prefer values $\approx 0.6$.

Regardless of the simulation, the bottom panel of Fig.~\ref{fig:barfraction} shows that the baryonic effects on the reduced equilateral bispectrum correlate with the baryonic fraction: the smaller the baryon fraction, the larger the bispectrum enhancement. Remarkably, the prediction from our model, when fitted with a simple linear regression,
shows a trend as tight as the one found in the power spectrum ($1\%$).
To have an idea of the predictions from hydrodynamical simulations, we infer the baryon fractions from the power spectra measurements using the fitting function provided by \cite{vandaalen2019}, and combine them to the measurement of the reduced bispectra enhancement. By doing so, we find that all the predictions are still within $1\%$, except for BAHAMAS high-AGN, which is slightly off but still well within $2\%$.

It would be very interesting to extend the analysis of \cite{vandaalen2019} to the bispectrum, to check if including a vast number of hydrodynamical simulations the relation still holds with a low scatter. Nevertheless, we stress that, a priori, the BCM does not predict a tight relation between baryon fraction and clustering. In A20 (Fig. 8) it was shown that the baryon fraction-clustering relation is more relaxed when considering the full BCM parameters ranges. Interestingly, it seems that the calibration and subgrid physics with which hydrodynamical simulations are run, translates into constraints and degeneracies of the BCM parameters, and thus constrain the baryon fraction.

\section{Conclusions}
\label{sec:conclusions}

In this paper, we have used a combination of cosmology scaling and baryonification algorithms, to reproduce with a  negligible computational time the density fields of various hydrodynamical simulations, up to very small and non-linear scales ($k=5\ihMpc$) and for two- and three-point statistics.

Below we summarise our main findings:

\begin{itemize}
\item Baryonic physics causes an enhancement in the reduced equilateral bispectrum at all the scales considered, roughly monotonically with the strength of the feedback mechanisms;

\item It is possible to simultaneously reproduce the baryonic effects on the power spectrum and on reduced bispectrum (with $1\%$ and $2-3\%$ precision, respectively), as measured in EAGLE, Illustris, Illustris TNG-300, and three different AGN implementations of BAHAMAS,

\item In contrast, a baryon model tuned to only reproduce the power spectrum, can lead to up to $\sim20\%$ discrepancies in the reduced bispectrum;

\item We find that a double power-law gas density profile is flexible enough to reproduce the bump at small scales measured in the bispectrum of some hydrodynamical simulations \cite[see][]{Foreman2019}. It appears thus that an additional modelling of gas overdensity at relatively small scales is superfluous.

\item The model parameters that best fit the power spectrum and equilateral bispectrum also predict changes to the squeezed configurations at the $\sim 1\%$ level.

\item The baryon parameters are not redshift independent; ignoring their time dependence results in a $5\%$ inaccuracy in the power spectrum, and $10-20\%$ in the reduced bispectrum, up to $z=2$;

\item Analysing the best-fitting models to the hydrodynamical simulations, we find a correlation between baryonic effects on the bispectrum and baryon fraction inside haloes, similar to the one for the power spectrum found in \cite{vandaalen2019},

\end{itemize}

Overall, our results support the physical soundness (as well as our specific numerical implementation) of baryonification algorithms. This also encourages its use not only in spherically-averaged 2-point statistics, but also in cross-correlations and in other statistics such as peak counts.

The next generation surveys will produce a huge amount of data, which is only partially interpretable with the current theoretical models. This paper is a contribution to the effort to overpass models based on only gravitational interactions, and fully exploit the data up to higher-order statistics. We anticipate that our approach will be a valid tool for a fast production of mock density fields, accurate to very small scales and statistics of order higher than 2-point, useful for pipeline validation, blind comparisons or for direct exploitation of the data, e.g. marginalising over baryonic effects.

\section*{Acknowledgements}

The authors acknowledge the support of the E.R.C. grant 716151 (BACCO). C.H.-M. acknowledges support from the Spanish Ministry of Economy and Competitiveness (MINECO) through the projects AYA2015-66211-C2-2 and PGC2018-097585-B-C21. SC acknowledges the support of the ``Juan de la Cierva Formaci\'on'' fellowship (FJCI-2017-33816). We thank Simon Foreman
for making public the power spectra and bispectra measurements used in this work, as well as the code {\bf bskit}.
We acknowledge the Illustris, Illustris TNG, BAHAMAS, and EAGLE teams, for providing/making public the data
of their hydrodynamical simulations.
We are grateful to Alex Barreira and Simon Foreman for carefully reading the draft, and helping us to improve the manuscript with their precious feedback.
We thank Yetli Rosas-Guevara for providing us with the measurements of the mass density profiles in the Illustris TNG-300 simulation.
G.A. thank Lurdes Ondaro Mallea and Marcos Pellejero Iba\~{n}ez for useful discussions.

\section*{Data Availability}

The data underlying this article will be shared on reasonable request to the corresponding author.

\bibliographystyle{mnras}
\bibliography{bibliography} 



\appendix
\section{Convergence test}
\label{app:convergence}

\begin{figure}
  \includegraphics[width=.99\linewidth]{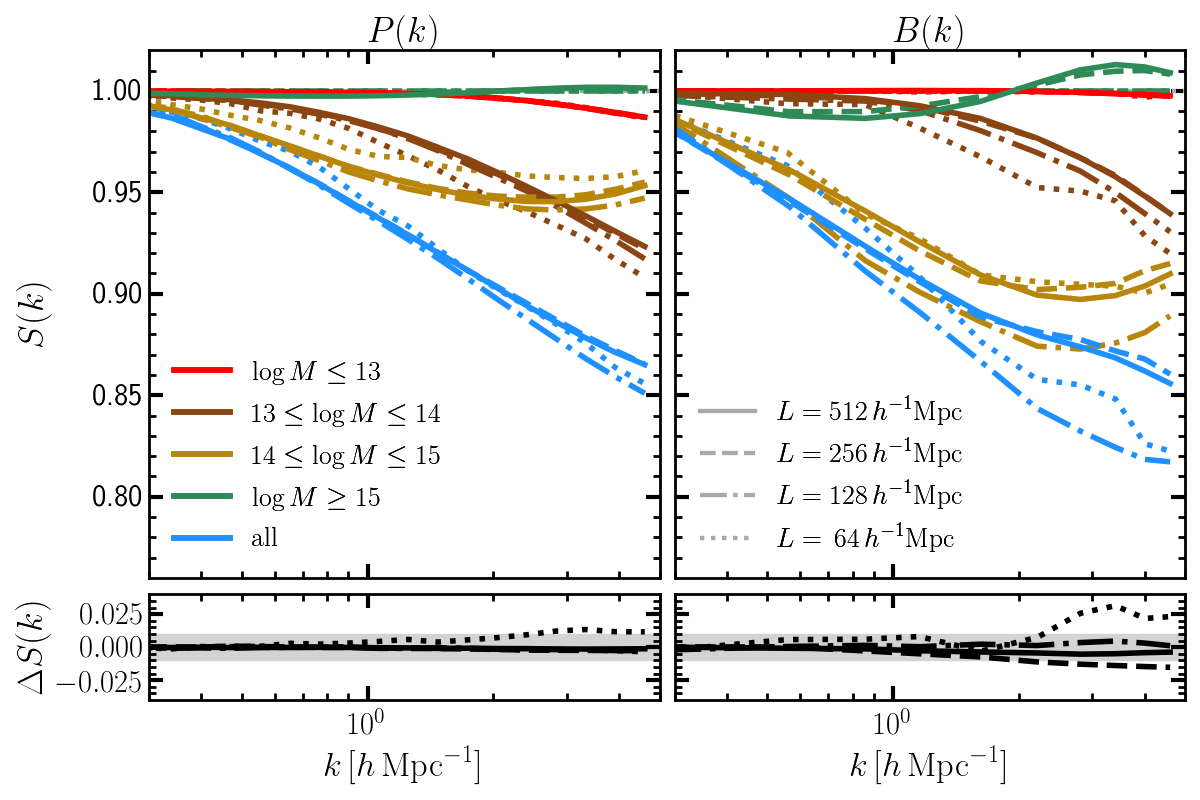}
  \caption{ {\it Upper panel:} Baryon suppression of the matter power spectrum (left) and reduced bispectrum (right), defined as $S(k) \equiv T(k)_{\rm BCM}/T(k)_{\rm GrO}$ for $T(k)=P(k),Q(k)$, at $z=0$. Solid, dashed, dashed-dotted and dotted lines
          are computed with simulations of box side  $512$, $256$, $128$ and $64 \, \hMpc$ and 1536$^3$, 768$^3$, 384$^3$ and 192$^3$ particles, respectively.
        Colors are referred to different halo mass bins, expressed in decimal logarithm of $\Msun$, with which the baryon corrections  have been computed, according to the legend.
          {\it Lower panel:} Difference in suppression between a {\it paired and fixed} simulation and a single realisation, for the four different volumes specified in the legend of the upper panel. }
          \label{fig:convergence}
  \end{figure}
  \begin{figure}
  \includegraphics[width=.99\linewidth]{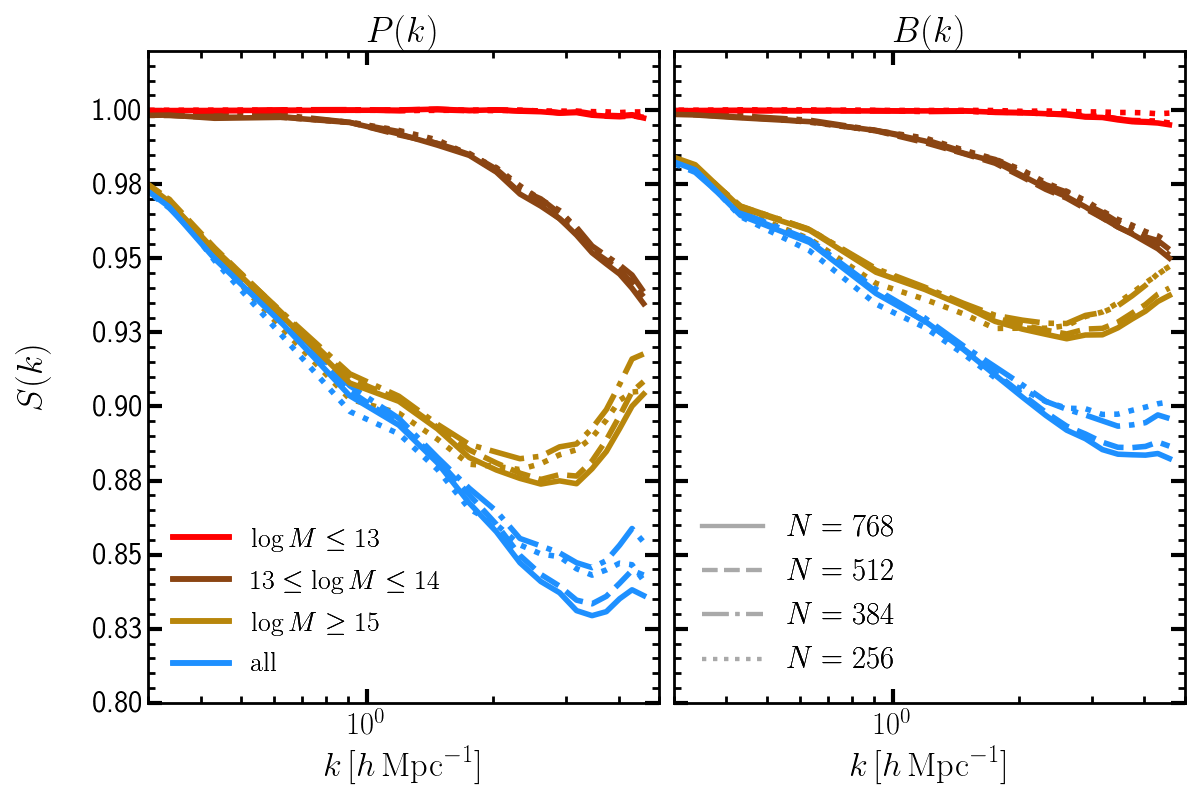}
  \caption{ Baryon suppression of the matter power spectrum (left) and reduced bispectrum (right), defined as $S(k) \equiv T(k)_{\rm BCM}/T(k)_{\rm GrO}$ for $T(k)=\{P(k),Q(k)\}$, at $z=0$. Solid, dashed, dashed-dotted and dotted lines are computed with simulations of box side  $133 \, \hMpc$ and 768$^3$, 512$^3$, 384$^3$ and 256$^3$ particles, respectively.
    Colours are referred to different halo mass bins, expressed in decimal logarithm of $\Msun$, with which the baryon corrections  have been computed, according to the legend.}
          \label{fig:resolution}
\end{figure}

In this Appendix, we show the tests we have performed to assure that the baryonic effects on the clustering measurements have converged.

First, we test the convergence with the simulation box size. For this, we have used our suite of simulations with $64 \, \hMpc$, $128 \, \hMpc$,  $256 \, \hMpc$, and  $512 \, \hMpc$ of box side, with $N=192$, $N=384$, $N=768$ and $N=1536$ cubic particles, respectively. All the simulations have same force and mass resolution, and share the same initial conditions. For each different volume we have run two simulations with fixed amplitude and shifted phases as reported in \S\ref{sec:sim}.

In Fig.~\ref{fig:convergence} we show the suppression $S(k)$, defined as the ratio between baryonified and gravity-only matter power spectra and equilateral bispectra, measured in the four different boxes at $z=0$.

We have also separated the contribution to the clustering of different halo masses, to get more insight on the origin of the discrepancies between the different boxes. As expected, we note that the boxsize does not affect sensibly haloes of $M\le10^{14}\, \Msun$. However, the abundance of massive haloes ($M=10^{14}-10^{15} \, \Msun$) varies consistently among the various boxes, leading to discrepancies in the baryonic effects that are still within $1\%$ in the power spectrum, but slightly higher in the bispectrum ($3-4\%$).

In the bottom panels of Fig.~\ref{fig:convergence} we display the impact of using a {\it paired and fixed simulation} against a single realisation. Also in this case, the biggest impact is found in the bispectrum, with a maximum of $\approx2.5\%$ bias when using a $64 \, \hMpc$ box, whereas we detect a maximum of $\approx1\%$ in the power spectrum. In the analysis, we make use of a single realisation of the $256 \, \hMpc$ box, which is shown to be converged within $2\%$.

We have performed also a mass resolution test, by using four simulation with the same box size, $133 \, \hMpc$, and different number of particles: $N=256^3$, $N=384^3$, $N=512^3$, $N=768^3$ particles. Also in this case, we split the contribution of different halo mass bins. As shown in Fig.~\ref{fig:resolution}, we have found that resolution effects are larger in large haloes, and their impact in the power spectrum and bispectrum is within $\approx 2\%$.

\begin{figure}
  \includegraphics[width=\linewidth]{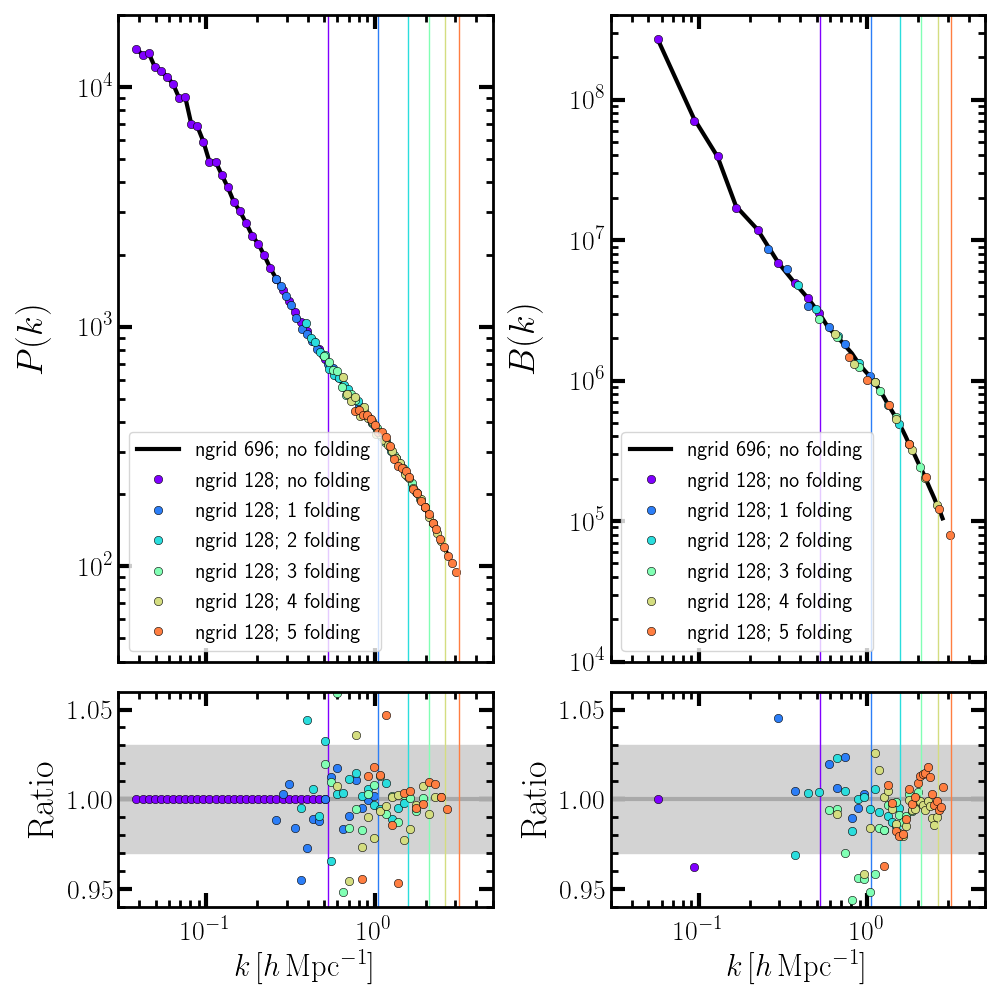}
  \caption{ {\it Left panels:} Matter power spectrum measured in a $64^3$ (interlaced) mesh, folding the box up to 6 times, following the technique explained in the text (coloured dots). For comparison,   the matter power spectrum measured with a $696^3$ mesh and not folding the box is plotted as a black solid lines. The equivalent Nyquist frequencies for each folded box is plotted as a  solid line. In the bottom panel, we display the ratio between the power spectrum measure with a $64^3$ and a $696^3$ mesh.
  {\it Right panels:} Similarly to the left panels, we display the measured bispectra using a $128^3$ mesh and the folding of the box, and compare with a not-folded $696^3$ mesh.
  }
\label{fig:folding}
\end{figure}

\begin{figure}
  \includegraphics[width=\linewidth]{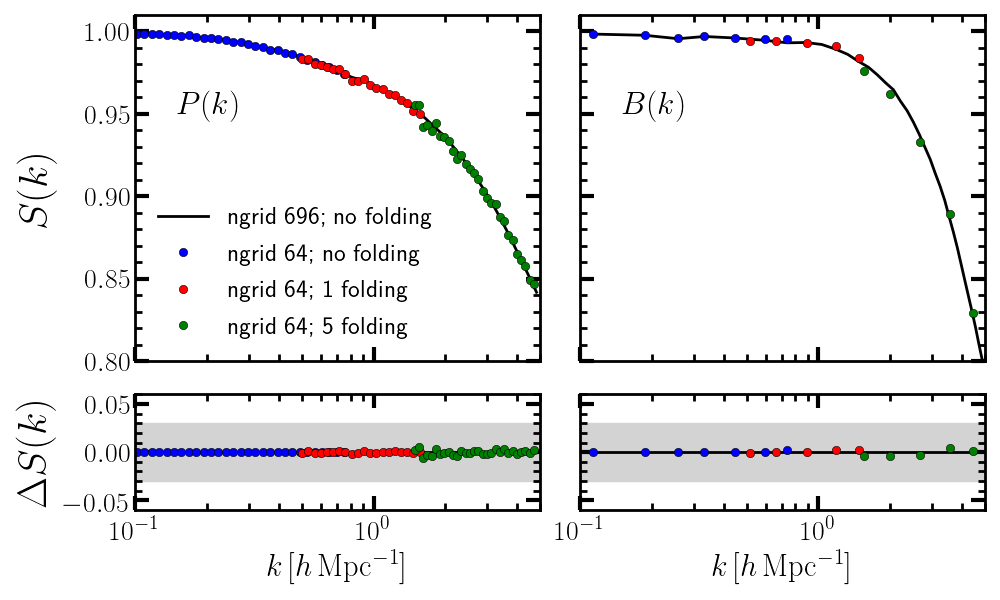}
  \caption{ {\it Left panels:} Ratio of baryonified and GrO matter power spectra $S(k)$, measured in a $64^3$ (interlaced) mesh, folding the particles up to 6 times, following the technique explained in the text (coloured dots). For comparison,
  the measurements with a $696^3$ mesh is plotted as a black solid lines. The equivalent Nyquist frequencies for each folded box is plotted as a coloured solid line. In the bottom panel, we display the difference of the ratios $\Delta S(k)$ measured with a $64^3$ and a $696^3$ mesh.
  {\it Right panels:} Similar to the left panels, but displaying the matter bispectrum instead of the power spectrum.
  }
\label{fig:folding_bcm}
\end{figure}

\section{Folding of the particle distribution}
\label{app:fold}

Measuring the three-point clustering with the classical Fourier estimators can be very expensive in terms of memory and CPU, especially when using covering larger dynamical ranges. In fact, it is easy to see that, being $k_{\rm Ny}= \pi N_g/L_{\rm box}$ the Nyquist frequency of the grid, increasingly large number of grid points $N_g$ are required
to get a given accuracy at a fixed wavenumber, when using progressively larger simulation boxes $L_{\rm box}$. Additionally, when using ``interlacing'' to suppress aliasing, the number of grids used must be doubled \citep{Sefusatti2016}.

However, since our measurements are limited by discreteness noise (and not cosmic variance), we can obtain accurate estimates of Fourier statistics on small scales by folding the density field \citep{Jenkins1998,Colombi2009}. The idea is to fold the particle distribution by re-applying the periodic boundary conditions assuming a new boxsize $L'=L/f$, where we call $f$ the number of foldings.
If $L'$ is large enough to assure that the modes inside the new box are uncorrelated, we can measure in principle the clustering from a new effective fundamental wavenumber $k_{\rm f}'= 2 \pi /L'$ up to a new effective Nyquist frequency, given by $k_{\rm Ny}'= \pi N_g/L'$. For instance, by folding the box 4 times, we will get to Nyquist frequency 4 times higher.



In Fig.~\ref{fig:folding} we apply this technique to our $512 \hMpc$ simulation, folding the particles up to 6 times, and reaching a $k_{\rm Ny} \approx 2 \ihMpc$ with a $64^3$ and a $128^3$ grid for the power spectrum and the bispectrum, respectively. Even using a TSC scheme on interlaced grids, we note that it is safer to use the measurements up to $k=k_{Ny}'/2$ in the bispectrum. Also, the measurement of the largest modes of the folded box are noisy because they are sparsely sampled; for this reason, it is convenient to discard these modes, taking for instance wavenumbers $k>10 k_{\rm f}'$.

Using these precautions, we show in Fig.~\ref{fig:folding_bcm} that using this technique
we can achieve an accuracy well within $1\%$ in the estimation of the ratios, using a
small fraction $1-10\%$ of the computational resources. Although it is common to use the folding technique to compute power spectra, to our knowledge, this is the first time it has been shown to be accurate for bispectrum measurements.

\bsp	
\label{lastpage}
\end{document}